\documentclass[aps,showpacs,preprintnumbers,amsmath, amssymb]{revtex4}

\usepackage{float}
\usepackage{graphics,epsfig}
\usepackage{graphicx}
\usepackage{dcolumn}
\usepackage{bm}

\begin{document}
\baselineskip=0.8 cm

\title{{\bf Studies of a general flat space/boson star transition model in a box through a language similar to holographic superconductors}}
\author{Yan Peng$^{1}$\footnote{yanpengphy@163.com}}
\affiliation{\\$^{1}$ School of Mathematical Sciences, Qufu Normal University, Qufu, Shandong 273165, China}

\vspace*{0.2cm}
\begin{abstract}
\baselineskip=0.6 cm
\begin{center}
{\bf Abstract}
\end{center}

We study a general flat space/boson star transition model
in quasi-local ensemble through approaches familiar from holographic superconductor theories.
We manage to find a parameter $\psi_{2}$, which is proved to be
useful in disclosing properties of phase transitions.
In this work, we explore effects of the scalar mass, scalar charge and St$\ddot{u}$ckelberg mechanism on the critical
phase transition points and the order of transitions mainly from behaviors of the parameter $\psi_{2}$.
We mention that properties of transitions in quasi-local gravity are strikingly similar to
those in holographic superconductor models.
We also obtain an analytical relation $\psi_{2}\varpropto(\mu-\mu_{c})^{1/2}$,
which also holds for the condensed scalar operator in the holographic insulator/superconductor system
in accordance with mean field theories.

\end{abstract}

\maketitle
\newpage
\vspace*{0.2cm}

\section{Introduction}

The schwarzschild black holes in flat space with negative specific
heat cannot be in equilibrium with the thermal environment since there
is inevitable Hawking radiation.
For this case, York and other authors provided a way to bypass this problem by putting
the black hole in a box to construct a quasi-local thermodynamic ensemble \cite{York,Braden}.
In contrast, the AdS black holes are usually thermodynamically stable since in a certain sense the AdS boundary plays a role
of the box condition \cite{Hawking}. With the interest of the AdS/CFT
correspondence \cite{Maldacena,S.S.Gubser,E.Witten}, the AdS gravity has attracted
a lot of attentions, such as there are many literatures about holographic superconductors constructed in the AdS spacetime \cite{S. A. Hartnoll-1}-\cite{Yi Ling}.

In the Einstein-Maxwell theory, it was shown that phase structures of the system
in a box is similar to those of the AdS gravity \cite{Robert,P. Hut,Gibbons}. Then, it is very interesting
to include a scalar field to further compare the similarity between transitions
in a box and those of the holographic superconductor system.
Recently, P. Basu, C. Krishnan and P.N.B. Subramanian initiated a thermodynamic study of such systems constructed by a charged scalar field
coupled to electromagnetic field in asymptotic flat space with box boundary conditions \cite{Pallab Basu}.
Besides the flat space and normal RN black hole, it also admits hairy solutions of boson star and hairy RN black hole in this model.
These two types of hairy solutions are thermodynamically stable for certain region of parameters.
Firstly, it provides a way to evade the flat space no-hair theorems.
The more interesting result is that the overall phase structure of the system in a box is strikingly similar to
that of holographic superconductors in AdS gravity background.
For example, there is flat space/black hole transitions corresponding to the classical Hawking-Page transition in AdS gravity \cite{Hawking}.
And this model also admits black hole/hairy black hole transitions similar to those in holographic theories.
In addition, there are flat space/boson star transitions corresponding to holographic insulator/superconductor
and global AdS/boson star systems in AdS gravity \cite{Basu,Gary T.Horowitz-2}.
As a further step along this line, it will be very interesting to study
phase transitions in a box with condensation diagrams from asymptotic behaviors of the scalar field on the boundary,
which is in a language more familiar from holographic superconductors in AdS gravity.

On the other side, it is meaningful to generalize the flat space/boson star model in \cite{Pallab Basu} by including
a non-zero scalar mass and examine how the scalar mass will affect the critical phase transition
points, which has been studied in holographic models \cite{Gary T.Horowitz-3,Q. Pan,Yan Peng-1}.
Since there is only second order flat space/boson star phase transition in \cite{Pallab Basu} for large charge of the scalar field,
it is interesting to extend this discussion to smaller charge as it has been
proved that there is first order holographic insulator/superconductor phase transitions for very small charge \cite{Gary T.Horowitz-2}.
As is well known in \cite{S. Franco-1,S. Franco-2,Q. Pan-1,Yan Peng-2,R.-G. Cai,Yan Peng-3,Yan Peng-4},
St$\ddot{u}$ckelberg mechanism usually triggers first order holographic phase transitions,
so it is meaningful to examine whether
the St$\ddot{u}$ckelberg mechanism could trigger first order phase transitions in quasi-local gravity.
From another aspect, hairy black holes and boson stars in a box
were constructed and dynamical properties of phases have been investigated in \cite{Dolan,Supakchai}.
In addition, it was shown in \cite{Nicolas} that hairy black holes really represent the end-point of the
superradiant instability of RN black holes in a box.
So it is also meaningful to study properties of gravity systems
in a box from the thermodynamic aspects.

The next sections are planed as follows. In section II,
we construct the general flat space/boson star phase transition model in a box beyond the probe limit.
In section III, we manage to find a simple way to describe phase transitions
and observe various types of transitions by choosing different
model parameters of the scalar mass, scalar charge and
St$\ddot{u}$ckelberg mechanism.
We summarize our main results in the last section.

\section{Equations of motion and boundary conditions}

We begin with the transition model constructed by a charged scalar field
coupled to a Maxwell field in the background of four dimensional asymptotic flat
spacetime with a fixed radial coordinate $r=r_{b}$ as the time-like box boundary.
And the general Lagrange density in St$\ddot{u}$ckelberg form reads \cite{Pallab Basu,S. Franco-1,Yan Peng-2}:
\begin{eqnarray}\label{lagrange-1}
\mathcal{L}=R-F^{MN}F_{MN}-(\partial \psi)^{2}-F(\psi)(\partial \theta
-q A_{\mu})^{2}-m^{2}\psi^{2},
\end{eqnarray}

where q and $m$ are the charge and mass of the scalar field respectively.
$A_{M}$ stands for the ordinary Maxwell field and $F(\psi)$ is an arbitrary function of $\psi$.
In this work, we will adopt the form that $F(\psi)=\psi^2+q^2C_{4}\psi^{4}$ in the following calculation \cite{Yan Peng-2}.
$C_{4}=0$ corresponds to the usual transition model in Part A of section III and
when $C_{4}\neq 0$, we examine effects of St$\ddot{u}$ckelberg mechanism on transitions in Part B of section III.
Using the gauge symmetry $A_{\mu}\rightarrow A_{\mu}+\partial \alpha, \theta\rightarrow \theta+\alpha$,
we can set $\theta=0$ without loss of generality.

For simplicity, we study matter fields with only radial dependence in the form

\begin{eqnarray}\label{symmetryBH}
A=\phi(r)dt,~~~~~~~~\psi=\psi(r).
\end{eqnarray}

Considering the matter fields' backreaction
on the metric, we take the deformed four dimensional boson star solution as \cite{Pallab Basu}
\begin{eqnarray}\label{AdSBH}
ds^{2}&=&-g(r)h(r)dt^{2}+\frac{dr^{2}}{g(r)}+r^{2}(d\theta^{2}+sin^{2}\theta d\varphi^{2}).
\end{eqnarray}

From above assumptions, we obtain equations of motion as

\begin{eqnarray}\label{BHpsi}
\frac{1}{2}\psi'(r)^{2}+\frac{g'(r)}{rg(r)}+\frac{q^2F(\psi)\phi(r)^2}{2g(r)^2h(r)}+\frac{\phi'(r)^2}{g(r)h(r)}-\frac{1}{r^2g(r)}+\frac{1}{r^2}+\frac{m^2}{2g}\psi^2=0,
\end{eqnarray}
\begin{eqnarray}\label{BHpsi}
h'(r)-rh(r)\psi'(r)^2-\frac{q^2rF(\psi)\phi(r)^2}{g(r)^2}=0,
\end{eqnarray}
\begin{eqnarray}\label{BHphi}
\phi''+\frac{2\phi'(r)}{r}-\frac{h'(r)\phi'(r)}{2h(r)}-\frac{q^2F(\psi)\phi(r)}{2g(r)}=0,
\end{eqnarray}
\begin{eqnarray}\label{BHg}
\psi''(r)+\frac{g'(r)\psi'(r)}{g(r)}+\frac{h'(r)\psi'(r)}{2h(r)}+\frac{2\psi'(r)}{r}+\frac{q^2F'(\psi)\phi(r)^2}{2g(r)^2h(r)}-\frac{m^2}{g}\psi=0,
\end{eqnarray}
Where $F'(\psi)=\frac{d F(\psi)}{d \psi}=2 \psi+ 4 q^2 C_{4} \psi^{3}$. These equations are nonlinear and coupled, so we have to use the
shooting method to integrate the equations from $r=0$ to box boundary $r=r_{b}$ to search for the numerical solutions satisfying boundary conditions.
Around $r=0$, the solutions can be expanded as \cite{Pallab Basu}
\begin{eqnarray}\label{InfBH}
&&\psi(r)=a+b r^2+\cdots,\nonumber\\
&&\phi(r)=aa+bbr^2+\cdots,\nonumber\\
&&g(r)=1+A r^2+\cdots,\nonumber\\
&&h(r)=AA+BB r^2+\cdots,
\end{eqnarray}
where the dots denote higher order terms.
Putting these expansions into equations of motion and considering leading
terms, we have three independent parameters
$a$, $aa$ and $AA$ left to describe the solutions.
With the rescaling $r\rightarrow ar$, we could set $r_{b}=1$. Near the box boundary $(r=1)$, the asymptotic
behaviors of the matter fields are
\begin{eqnarray}\label{InfBH}
\psi\rightarrow \psi_{1}+\psi_{2}(1-r)+\cdots,\\
\phi\rightarrow \phi_{1}+\phi_{2}(1-r)+\cdots,
\end{eqnarray}
with $\mu=\phi(1)=\phi_{1}$ as the critical chemical potential. We will also make a transformation
to take $g_{tt}(1)=1$ with the symmetry $h\rightarrow a^2h,~\phi\rightarrow \phi,~t\rightarrow\frac{t}{a}$ \cite{Pallab Basu}.
Since we impose a mirror boundary conditions for the scalar field as $\psi(r_{b})=0$, we have to set $\psi_{1}=0$ and try to use
$\psi_{2}$ to describe the phase transition, which is similar to approaches in holographic superconductor theories.
We will show in the following section that $\psi_{2}$
is a good probe to properties of phase transitions
in quasi-local ensemble.
We also point out that our box boundary condition is independent
of the mass of scalar fields, which is different from that in holographic
superconductor theories where asymptotic behaviors of scalar fields at infinity boundary
usually depend on the mass.

\section{Properties of phase transitions in a box}

\subsection{Scalar condensation with various $q$ and $m^2$}

In this part, we firstly show the numerical boson star solutions in Fig. 1.
In the left panel, the scalar field decreases as approaching the box boundary and
at the boundary there is $\psi(r_{b})=0$. In the middle panel, the vector field
increases as a function of the radial coordinate. We also show behaviors of the
metric solutions $g(r)h(r)=-g_{tt}(r)$ in the right panel.
When neglecting the matter fields'
backreaction on the metric, we will have $g(r)h(r)=1$ and in contrast,
the curves in the right panel show that the metric is deformed by matter fields when including backreaction.

\begin{figure}[h]
\includegraphics[width=155pt]{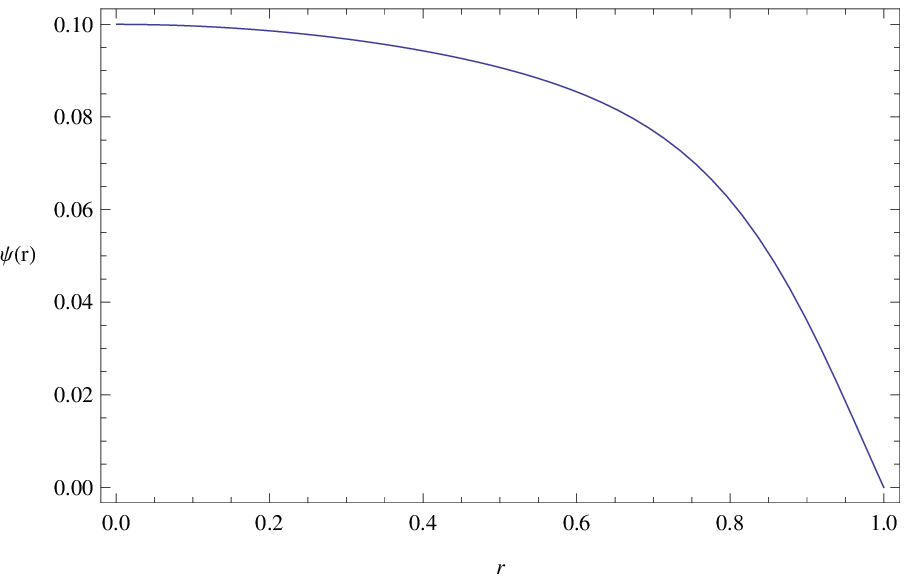}\
\includegraphics[width=155pt]{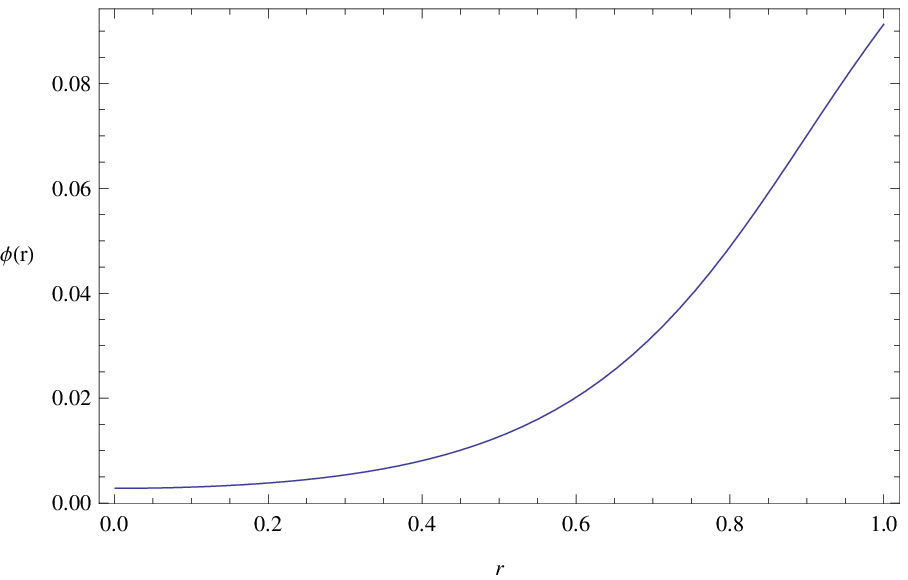}\
\includegraphics[width=155pt]{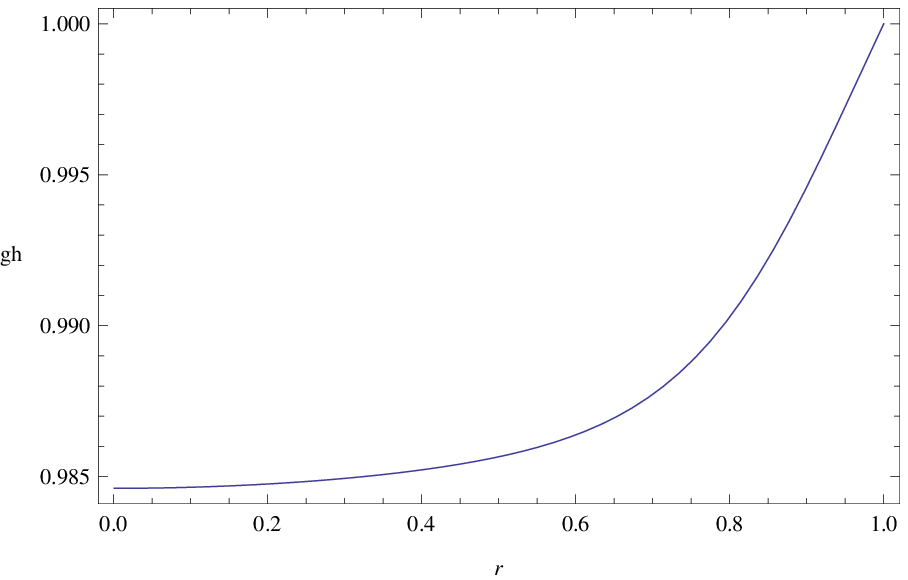}\
\caption{\label{EEntropySoliton} (Color online) We plot solutions as a function of the
radial coordinate $r$ with $q=100$, $m^{2}=-2$, $C_{4}=0$ and $\psi(0)=0.1$. The left panel shows behaviors
of $\psi(r)$, the middle panel corresponds to the vector field $\phi(r)$
and the right panel represents the values of $g(r)h(r)=-g_{tt}(r)$.}
\end{figure}

It is known that free energy can be used to determine the critical phase transition points and also the order of transitions.
The authors in \cite{Pallab Basu} have proposed a way to calculate
the free energy of the system by adding a Gibbons-Hawking-York term on the boundary of the box.
After subtracting the free energy of the flat space,
we arrive at a formula for the free energy of boson star expressed by the metric solutions
and the box boundary radial coordinate as \cite{Pallab Basu}

$F=\frac{1}{\sqrt{g(r_{b})h(r_{b})}}(-\frac{1}{2}\int_{0}^{r_{b}}\sqrt{h(r)}dr-\frac{r_{b}}{2}g(r_{b})\sqrt{h(r_{b})}
-\frac{r_{b}^{2}}{4}g'(r_{b})\sqrt{h(r_{b})}-\frac{r_{b}^{2}}{4}\frac{g(r_{b})}{\sqrt{h(r_{b})}}h'(r_{b}))-(-r_{b})$.

\begin{figure}[h]
\includegraphics[width=180pt]{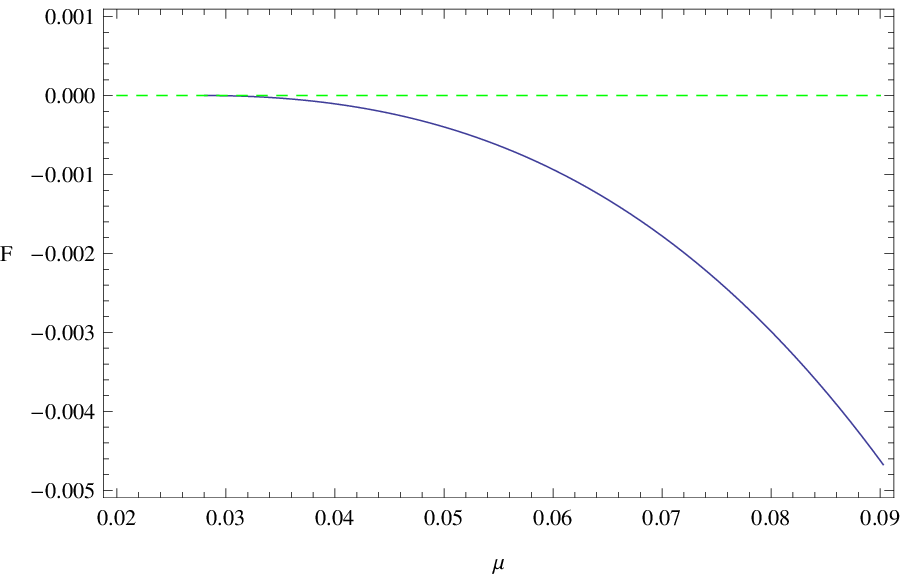}\
\includegraphics[width=180pt]{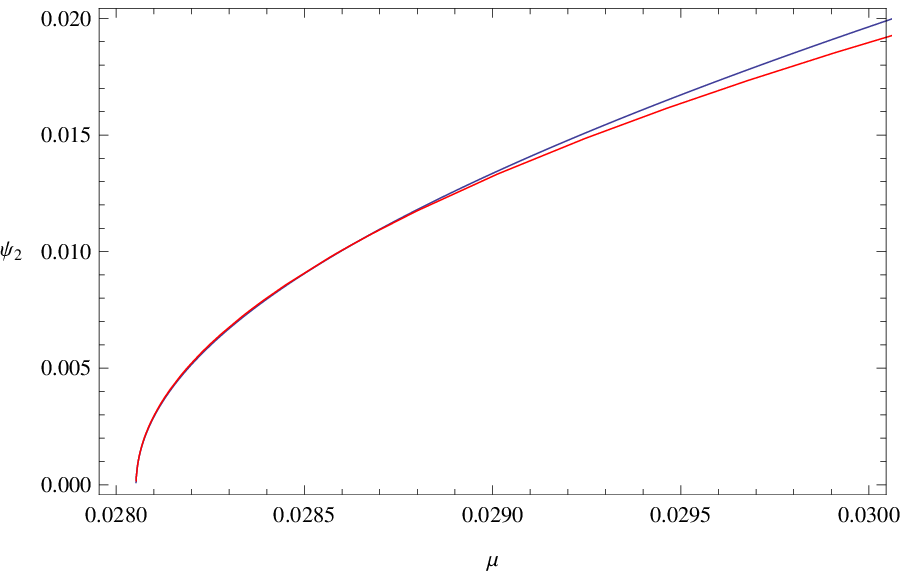}\
\caption{\label{EEntropySoliton} (Color online) We plot the free energy as a function of chemical potential
in the left panel with $q=100$, $m^{2}=-2$, $C_{4}=0$ and the right panel is with behaviors
of $\psi_{2}$. The solid blue line represents boson star phases and the solid red line corresponds to $\psi_{2}\thickapprox0.430(\mu-\mu_{c})^{1/2}$
with $\mu_{c}=0.02806$. In addition, the dashed line of $F=0$ in the left panel shows the free energy of the flat space.}
\end{figure}

We plot the free energy in the case with $q=100$, $m^{2}=-2$ and $C_{4}=0$ in the left panel of Fig. 2.
Here, we take $q=100$ in order to compare our data with results in \cite{Pallab Basu}.
For every fixed value of the chemical potential, we can choose only one phase.
Following the lowest free energy,
it can be seen from the left panel that the solid blue line
with a critical chemical potential $\mu_{c}=0.02806$ is physical.
Since the free energy is smooth as a function of the chemical potential around $\mu_{c}=0.02806$,
the flat space and boson star transition is of the second order.
With various charge q, we find that there is only second order phase transitions
for large charge $q> 4.5$, which is qualitatively the same to corresponding
properties in holographic insulator/superconductor systems \cite{Gary T.Horowitz-2}.

Inspired by the holographic superconductor theory, we also want to disclose properties
of transitions from condensation diagram directly related to the scalar field.
We choose to study $\psi_{2}$ as a function of $\mu$ in cases of $q=100$, $m^{2}=-2$
and $C_{4}=0$ in the right panel of Fig. 2 with solid blue line.
It can be seen from the right panel that $\psi_{2}$ increases as we choose
a larger chemical potential, which is qualitatively the same to properties in
the holographic insulator/superconductor transition \cite{T. Nishioka,Gary T.Horowitz-2}.
We also find a critical chemical potential $\mu=0.02806$ in the right panel equal to $\mu_{c}=0.02806$ in the left panel, above which the
parameter $\psi_{2}$ becomes nonzero. This implies that the parameter $\psi_{2}$ can be used to
determine the critical chemical potential of the transition.

By fitting the numerical data, we also obtain an analytical relation $\psi_{2}\varpropto (\mu-\mu_{c})^{1/2}$,
which also holds in the holographic insulator/superconductor transitions in accordance with
mean fields theories \cite{Rong-Gen Cai}. We have plotted the fitting formula $\psi_{2}\thickapprox0.430(\mu-\mu_{c})^{1/2}$ with $\mu_{c}=0.02806$
in the right panel of Fig. 2 with red solid line. It is clearly that the red solid line almost coincides
with the $\psi_{2}\thicksim \mu$ diagram in solid blue line around the phase transition points.
It seems that the parameter $\psi_{2}$ plays a role strikingly similar to
the scalar operator in holographic insulator/superconductor theories.

\begin{figure}[h]
\includegraphics[width=180pt]{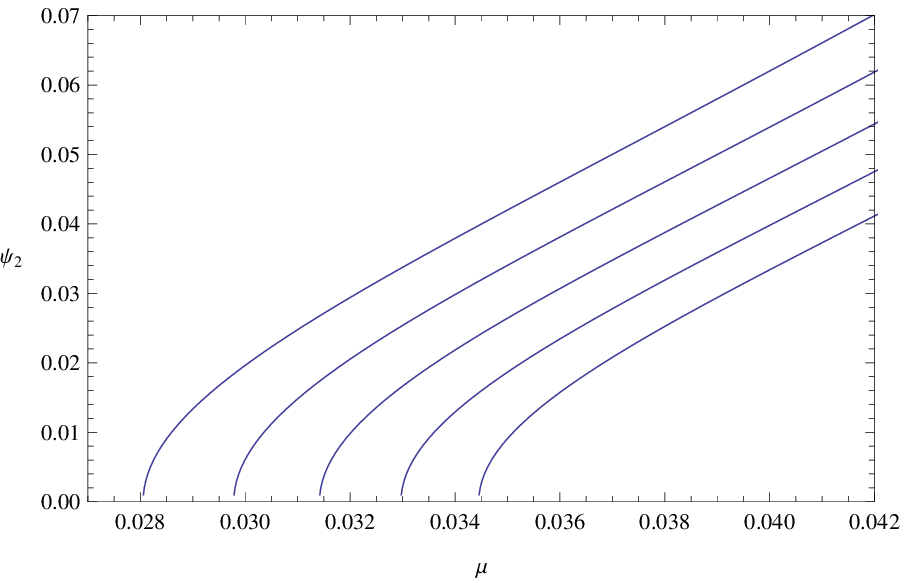}\
\includegraphics[width=180pt]{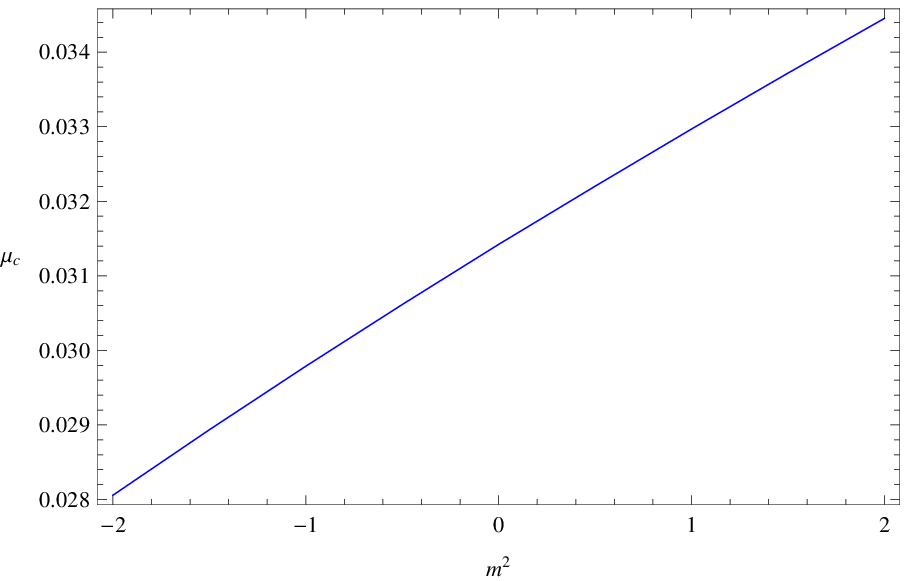}\
\caption{\label{EEntropySoliton} (Color online)
In the left panel, we show $\psi_{2}$ as a functions of $\mu$ with $q=100$, $C_{4}=0$
and various mass $m^2$ from left to right as: $m^2=-2$, $m^2=-1$, $m^2=0$, $m^2=1$ and $m^2=2$.
In the right panel, we plot the critical chemical potential $\mu_{c}$ as a function
of the scalar mass $m^2$ with $q=100$ and $C_{4}=0$.}
\end{figure}

For every set of parameters, we obtain a critical chemical potential $\mu_{c}$,
above which the flat space gives way to the boson star phases.
By choosing $q=100$, $C_{4}=0$ and various scalar mass $m^2$, we disclose effects of the
scalar mass on the critical chemical potential $\mu_{c}$ in the left panel of Fig. 3.
It can be easily seen from the left panel that $\mu_{c}$ is larger as we choose a larger $m^2$.
With more calculations, we go on to plot the critical chemical potential $\mu_{c}$ as a function of $m^2$ in the right panel of Fig. 3.
It can be easily seen from the picture that $\mu_{c}$ is almost in linear with respect to the scalar mass $m^{2}$.
By fitting the data, we find $\mu_{c}\thickapprox\frac{\pi}{100}+ 0.001596*m^2$. When $m^2=0$, it returns to $\mu_{c}=\frac{\pi}{q}$
in cases of zero scalar mass \cite{Pallab Basu}.
That is to say $\mu_{c}$ increases as we choose a less negative
scalar mass or larger mass
makes the second order phase transitions more difficult to happen,
which is in qualitatively accordance with the corresponding properties
in holographic insulator/superconductor theories.

\begin{figure}[h]
\includegraphics[width=180pt]{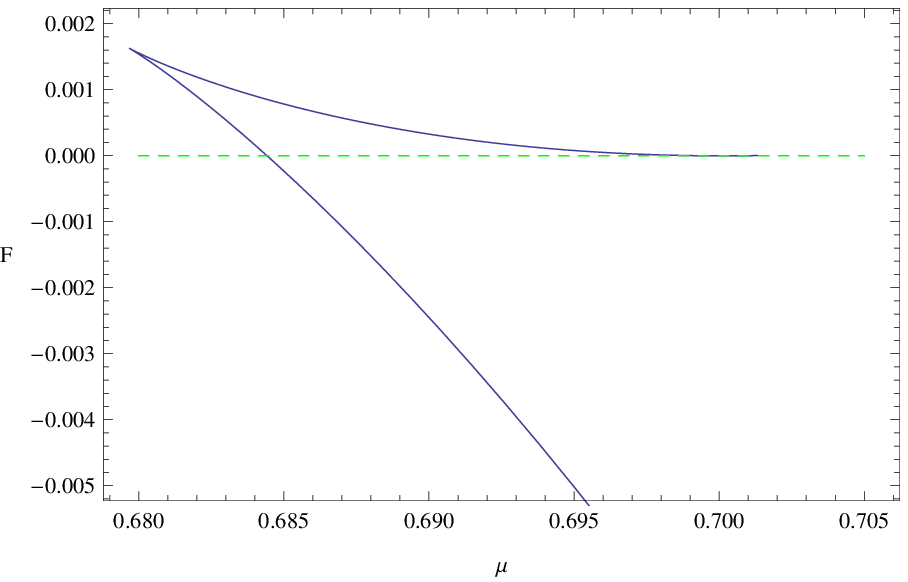}\
\includegraphics[width=180pt]{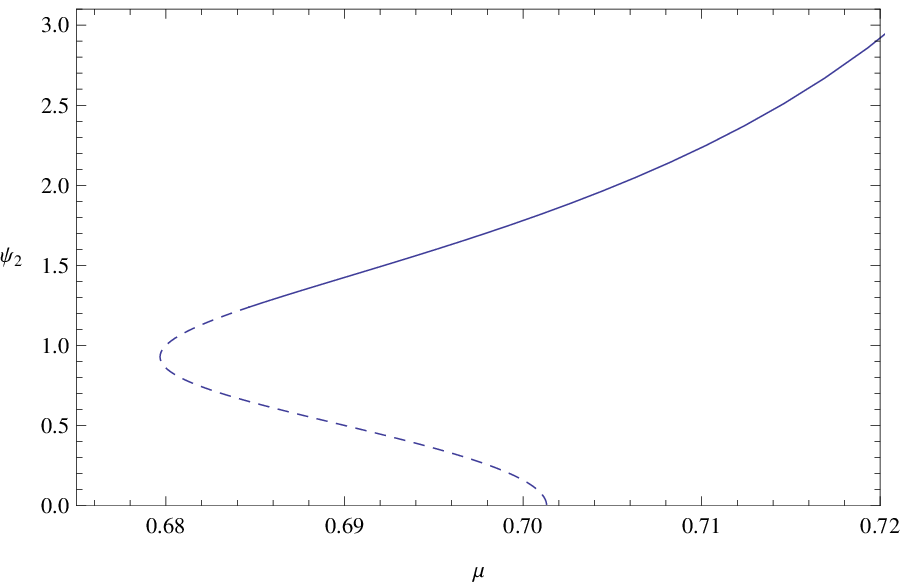}\
\caption{\label{EEntropySoliton} (Color online) We show phases with $q=4$, $m^2=-2$ and $C_{4}=0$.
In the left panel, we plot the free energy of boson star as a function of the chemical potential
with solid blue curves and the dashed green curves of $F=0$ is with the flat space.
The solid blue line in the right panel represents behaviors of parameter $\psi_{2}$ in
the thermodynamically stable phases and the dashed blue line in the right panel are with thermodynamically unstable phases.
}
\end{figure}

Now, we turn to study properties of phase transitions with small charge in our flat space/boson star transition model.
In the left panel of Fig. 4, in cases of $q=4$, $m^2=-2$ and $C_{4}=0$,
the free energy has a jump at the critical phase transition points $\mu_{c}=0.6844$
implying that there is a first order phase transitions.
We also show $\psi_{2}$ as a function of chemical potential in the right panel with $q=4$, $m^2=-2$ and $C_{4}=0$.
It can be seen that $\psi_{2}$ has a jump at the threshold value $\mu_{c}=0.6844$.
In other words, $\psi_{2}$ can be used to determine the threshold phase transition points
and also the order of transitions.
With more calculations, we conclude that small charge $q< 4.5$ could trigger first order flat space/boson
star transitions similar to cases in holographic insulator/superconductor system \cite{Gary T.Horowitz-2}.

\subsection{Phase transitions with St$\ddot{u}$ckelberg mechanism}

It is known that holographic superconductor models with St$\ddot{u}$ckelberg mechanism usually allows
first order transitions to occur for certain region of parameters in the background of
AdS gravity \cite{S. Franco-1,S. Franco-2,Q. Pan-1,Yan Peng-2,R.-G. Cai,Yan Peng-3,Yan Peng-4}.
Now we turn to examine whether the St$\ddot{u}$ckelberg mechanism
could trigger first order phase transitions for large charge, where there is only
second order transitions without St$\ddot{u}$ckelberg mechanism for $q> 4.5$ in Part A.
We show the free energy with respect to the chemical potential
$\mu$ in the left panel of Fig. 5 in cases of $q=100$, $m^2=-2$ and $C_{4}=0.1$.
Following the lowest line, we arrive at a critical chemical potential $\mu_{c}=0.02695$,
around which there is a jump of the slop of the free energy as a function of $\mu$.
That implies a first order flat space/boson star phase transition at $\mu_{c}$.
We also study the behaviors of $\psi_{2}$ with $q=100$, $m^2=-2$ and $C_{4}=0.1$ in the right panel
of Fig. 5. At the critical chemical potential $\mu_{c}=0.02695$, we have a jump of $\psi_{2}$.
Compared with behaviors of the free energy, we again conclude that the jump of $\psi_{2}$ implies a first
order phase transition.
So we include that St$\ddot{u}$ckelberg mechanism could trigger first order phase transitions
in gravity systems in a box just as properties found in holographic AdS gravity systems.

\begin{figure}[h]
\includegraphics[width=180pt]{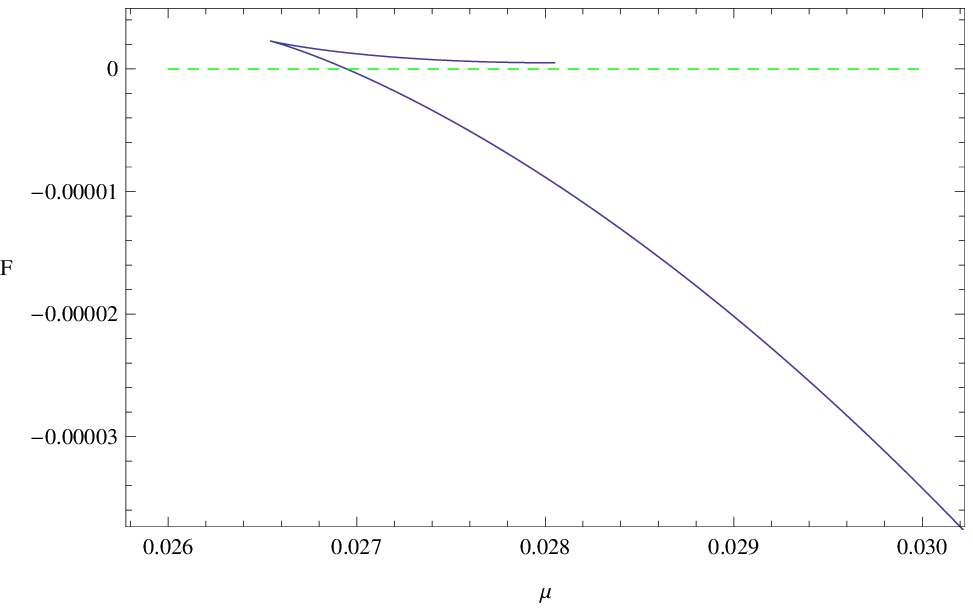}\
\includegraphics[width=180pt]{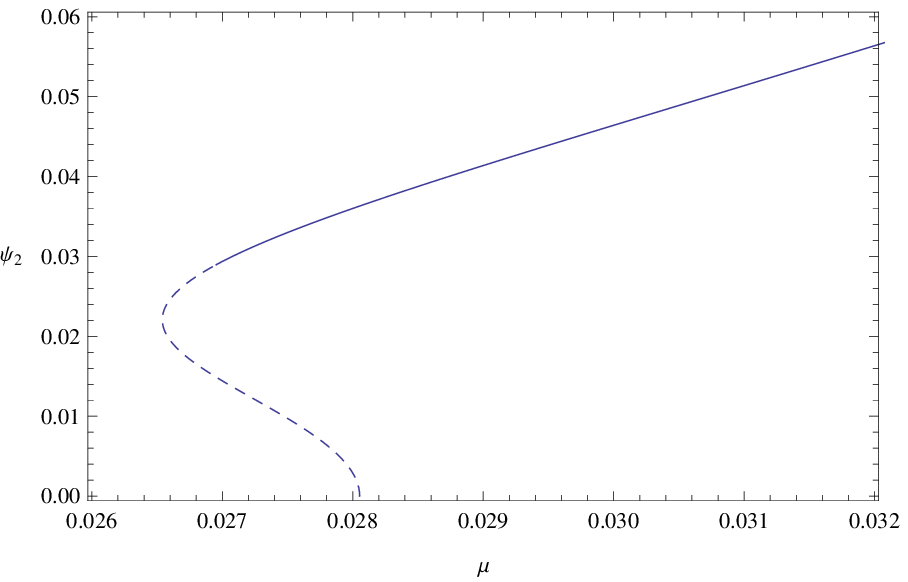}\
\caption{\label{EEntropySoliton} (Color online)
The two panels are with $q=100$, $m^2=-2$ and $C_{4}=0.1$.
We show behaviors of free energy of boson star phases with solid blue line in the left panel and
the dashed green line of $F=0$ in the left panel is with the flat space.
We also show behaviors of $\psi_{2}$ in the right panel.
In addition, the solid blue curve in the right panel represents
the stable phases and the dashed blue line in the right panel corresponds to unstable phases.
}
\end{figure}

Besides the parameter $\psi_{2}$, we also find that the metric solutions can be used to disclose
properties of phase transitions in our flat space/boson star transition model.
In the left panel of Fig. 6, when there are second order phase transitions with $q=100$, $m^2=-2$ and $C_{4}=0$,
$g(1)$ has a jump of the slope with respect to chemical potential at the critical phase transition
points $\mu_{c}=0.02806$. In contrast, for the case of first order phase transitions with
$q=100$, $m^2=-2$ and $C_{4}=0.1$ in the right panel of Fig. 6, $g(1)$ has a jump at the threshold value $\mu_{c}=0.02695$.
Compared with results in Fig. 2 and Fig. 5, we conclude that
metric solutions can be used to determine the threshold phase transition points
and the order of flat space/boson star phase transitions,
where similar results have been found in holographic metal/superconductor models \cite{Yan Peng-4}.

\begin{figure}[h]
\includegraphics[width=180pt]{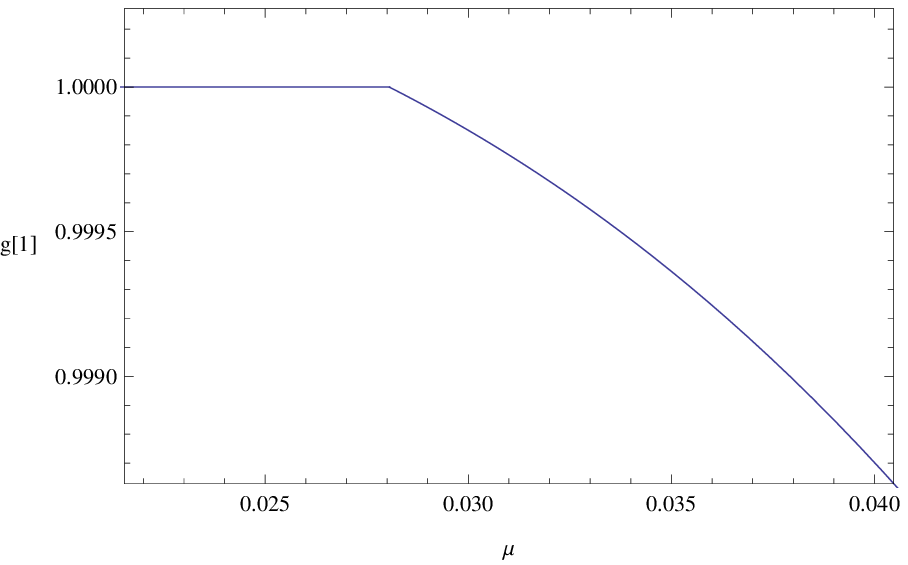}\
\includegraphics[width=180pt]{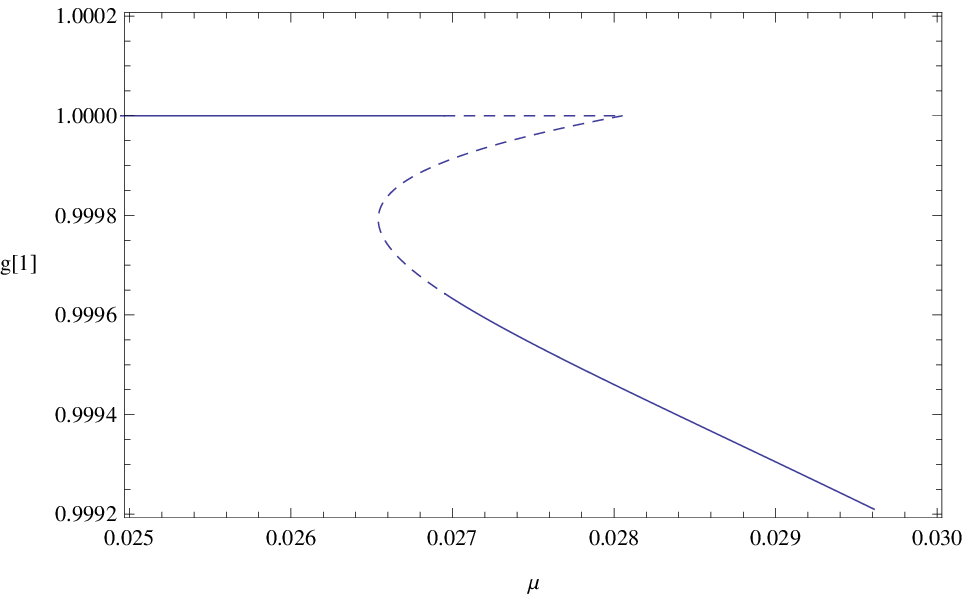}\
\caption{\label{EEntropySoliton} (Color online)
We show $g(1)$ with respect to chemical potential $\mu$.
In the left panel, the curve is with phases of $q=100$, $m^2=-2$ and $C_{4}=0$ and
the right panel represents cases of $q=4$, $m^2=-2$ and $C_{4}=0.1$.
We show stable phases with solid curves and unstable phases with dashed curves.}
\end{figure}

\section{Conclusions}

We studied a general four dimensional flat space/boson star transition model in a box beyond the probe limit.
With numerical methods, we deduced the parameter $\psi_{2}$ from behaviors of the scalar field on the box
boundary and found that $\psi_{2}$ can be used to detect the critical phase transition points
and also the order of transitions.
Similar to holographic insulator/superconductor transitions,
we found that for large scalar charge $q>4.5$, there is only second order transitions and small charge $q< 4.5$ could
trigger first order phase transitions.
For the second order transitions, we obtained an analytical relation $\psi_{2}\varpropto(\mu-\mu_{c})^{1/2}$,
which also holds in the holographic insulator/superconductor system in accordance with mean field theories.
As a summary, we examined effects of the scalar mass on the second order critical chemical potential
mainly from behaviors of $\psi_{2}$.
We found the more negative scalar mass makes the second order phase transition
more easier to happen qualitatively in accordance with cases in holographic insulator/superconductor models.
For large scalar charge $q>4.5$, St$\ddot{u}$ckelberg mechanism could make the transformation of the
order of phase transitions from the second order into the first order and the metric solutions
were also proved to be useful in detecting properties of transitions
which are similar to cases in holographic theories.
In summary, we obtained richer physics mainly through behaviors of $\psi_{2}$
in the generalized flat space/boson star transition model with box boundary conditions by
including non-zero scalar mass, various scalar charge
and St$\ddot{u}$ckelberg mechanism. And we also pointed out that
effects of scalar mass, scalar charge and St$\ddot{u}$ckelberg mechanism on transitions in a box is
qualitatively the same with those of holographic insulator/superconductor theories in AdS gravity
and our results provided additional hints of the
existence of similar holographic theories in gravity systems in a box.

\begin{acknowledgments}

We would like to thank the anonymous referee for the con-structive comments to improve the manuscript.
This work was supported by the National Natural Science Foundation of China under Grant No. 11305097;
the Shaanxi Province Science and Technology Department Foundation of China
under Grant No. 2016JQ1039.

\end{acknowledgments}

\end{document}